\documentclass[aps, preprint, showpacs]{revtex4}
\begin{document}
\title{Rotating light, OAM paradox and relativistic complex scalar field}
\author{S. C. Tiwari \\
Institute of Natural Philosophy \\
1 Kusum Kutir, Mahamanapuri \\
Varanasi 221005, India }
\begin{abstract}
Recent studies show that the angular momentum, both spin and orbital, of rotating light beams possesses counter-intuitive
characteristics. We present a new approach to the question of orbital angular momentum of light based on the
complex massless scalar field representation of light. The covariant equation for the scalar field is treated in rotating system using the general relativistic framework. First we show the equivalence of the U(1) gauge current for the scalar field with the Poynting vector continuity equation for paraxial light, and then apply the formalism to the calculation of the orbital angular momentum of rotating light beams. If the difference between the co-, contra-, and physical quantities is properly accounted for there does not result any paradox in the orbital angular momentum of rotating light. An artificial analogue of the paradoxical situation could be constructed but it is wrong within the present formalism. It is shown that the orbital angular momentum of rotating beam comprising of modes with opposite azimuthal indices corresponds to that of rigid rotation. A short review on the electromagnetism in noninertial systems is presented to motivate a fully covariant Maxwell field approach in rotating system to address the rotating light phenomenon. 
\end{abstract}
\maketitle


\section{\bf Introduction}

Rotating light beams seem to possess interesting and counter-intuitive properties:
frequency shift, angular momentum (AM) opposite to the direction of the beam rotation,
and frequency-dependent spin angular momentum (SAM). Recall that the rotation of the
fields is inherent in the electromagnetic waves \cite{1} , for example, in a circularly
polarized plane wave the electric field vector at a fixed point in space rotates at the
frequency $\omega$ of the wave. Rotating light beams, however correspond to the 'forced
rotation' \cite{2} employing rotating optical media: rotating half-wave plate for 
polarization rotation \cite{3} and rotating Dove prism (or a $\pi$ mode converter) for
mode pattern rotation of the paraxial modes of the Laguerre-Gaussian (LG) beams \cite{4,5}.
Until recently the main concern had been with the rotational frequency shifts. Since
polarization is associated with SAM, and the LG modes carry orbital angular momentum (OAM)
\cite{6} it is logical to investigate the effect of rotation on them. Straightforward
generalization of the theory of \cite{6} with the use of time-dependent paraxial
wave equation \cite{4} is obtained introducing uniform rotation frequency $\Omega$
replacing the azimuthal coordinate $\phi$ to $\phi -\Omega t$
in the mode function \cite{2}. The vector mode function for the case of the polarization and polychromatic waves are introduced in \cite{7}. The calculated OAM and SAM are found to be oppositely directed to the angular velocity; this is termed 'paradoxical' \cite{2} and 
counter-intuitive \cite{7}. Alexeyev and Yavorsky \cite{8} also note that the application
of Berry's formula \cite{9} leads to a 'confusing result', namely the $\Omega$- 
independent AM. The role of polychromatic nature of light for rotating beams is recognised
by all of them, and qualitative arguments based on the notion of photon are presented
seeking the resolution of the paradox. A recent work \cite{10} addresses the problem
developing quantum optics formalism for rotating light.

Though useful insights have been obtained in the cited literature, there do exist
gaps in the understanding of the AM of rotating light. In the present paper we envisage
a fresh approach based on the classical relativistic scalar field. It is surprising
that apparently unphysical result that SAM depends on frequency is not noticed in the
current literature. The fact that spin of photon is independent of frequency, played
an important role in the physical interpretation of the angular Doppler effect \cite{11}.
Further the Sagnac effect truely belongs to the class of phenomena in which rotation
causes frequency shifts. Therefore a brief appraisal of this effect in the context
of the controversies associated with rotating light is also presented.

The main contribution of the present work is to develop a relativistic scalar field
theory for rotating light. Note that treating a component of the electric field or
the vector potential, in some cases, as a complex scalar function is well known \cite{1};
typical laser modes of 'doughnut' shape having helical wavefronts (phase singularities)
are also describable as complex scalar (electric) field \cite{12}, and the polarization-
a typical vector property of light could be operationally defined based on the intensity
(a scalar function) measurements \cite{13}, see also \cite{14}. What differentiates
our approach is that a massless complex scalar field $\Psi$ without recourse to
the electromagnetic fields is shown to possess a nice property: exact equivalence of
the Noether conserved current corresponding to the gauge symmetry \cite{15} with the
Poynting vector calculated in the paraxial approximation, Eq.(2.5) of \cite{16}. Thus
OAM of not only LG modes but that of multipole fields \cite{1} follows immediately in
this formalism. This allows us to generalize the approach to the covariant form of
scalar wave equation in a rotating coordinate system a la general relativity \cite{17}.
Bichromatic field becomes imperative, and frequency shift follows from the dispersion 
relation. The expression for the Noether current gives the Poynting vector equivalent
without any ambiguity. This allows us to resolve the OAM paradox satisfactorily.

The paper is organized as follows. In the next section key results given in \cite{16}
are reproduced for a self- contained presentation, and some critical remarks are made concerning
AM of light. In Sec.III conserved current associated with the gauge invariance of
the complex scalar field is shown to be identical with the Poynting vector for
paraxial linearly polarized beam. It is argued that the application of this result to 
multipole fields indicates its general validity. The covariant form of scalar field
equation in a rotating frame is presented in Sec.IV. Assuming cylindrical symmetry
and azimuthal dependence of the form $exp (i l \phi)$ dispersion relation is derived.
Using the superposition of the scalar fields OAM is calculated from the Noether current.
It is remarkable that the OAM of rotating beam could be interpreted in terms of rigid rotation, and
there do not arise any counter-intuitive features. In Sec.V the question of OAM paradox is revisited, and
resolved. A brief review on the electromagnetic fields in noninertial systems, and its relevance
in the present context constitute Sec.VI. The limitations and the outlook of our approach are discussed
in the last section.

\section{\bf The paraxial approximation and optical angular momentum}

Paraxial rays in optics have been known since long. Paraxial wave equation describes
reasonably well the laser beam propagation \cite{19}. Assuming monochromatic plane
polarized wave propagating along z-axis the electric field can be represented in
terms of a scalar function $u(x,y,z)$ and a phase factor $e^{ikz-i\omega t}$. In
the paraxial approximation the second derivative of $u$ with respect to z is
neglected compared with $|k\frac {\partial u}{\partial z}|$ so that the wave equation
reduces to
\begin{equation}
\nabla _t ^2 u =-2 i k \frac {\partial u}{\partial z}
\end{equation}
Here $\nabla _t^2$ is 2-dimensional Laplacian in the transverse plane and the wave number
$k$ is given by $\omega =c k$ in free space. Solution of paraxial wave equation, Eq.(1)
is inconsistent with the Maxwell equation \cite{20}: for an x-polarized wave the electric
field must be independent of x, however the lowest mode is Gaussian in x and y. Lax
et al \cite{20} find it paradoxical: Experimentally the laser-oscillator modes found in this
apparently inconsistent way agree extremely well with those predicted by this thoery. A 
systematic procedure assuming power series expansion of the electromagnetic fields in terms
of the ratio of the beam size in transverse plane to the diffraction length in the longitudinal direction is developed by them. Allen et al \cite{6} following \cite{21} assume the paraxial
scalar wave equation for the vector potential instead of the electric field. Let
\begin{equation}
{\bf A} = \hat {\bf x} u e^{i(k z-\omega t)}
\end{equation}
and the Poynting vector is defined to be
\begin{equation}
{\bf S} =\epsilon _0 {\bf E}\times {\bf B}
\end{equation}
The time-averaged Poynting vector for the paraxial light is finally obtained to be
\begin{equation}
{\bf S}^{(1)} =\frac {i\omega \epsilon _0}{2} [u\nabla u^* -u^* \nabla u -2iku^*u 
\hat {\bf {z}}]
\end{equation}
Expression (4) is Eq.(2.5) of \cite{16} referred to above in the preceding section and
represents the linear momentum density in the beam.

The angular momentum density
\begin{equation}
{\bf j} = \epsilon _0 {\bf r}\times ({\bf E} \times {\bf B})
\end{equation}
can be calculated from Eq.(4). For a cylindrical beam profile having
\begin{equation}
u(r,\phi ,z)=u_0 (r,z) e^{il\phi}
\end{equation}
the z-component of the AM density is calculated to be
\begin{equation}
j_z^{(1)}= \epsilon _0 \omega l u^* u
\end{equation}
The energy density in this case is given by the product of the linear momentum density
and the velocity of light
\begin{equation}
w=c~\epsilon _0 \omega k u^*u
\end{equation}
Evidently
\begin{equation}
\frac {j_z^{(1)}}{w} =\frac {l}{\omega}
\end{equation}
Since the light beam does not carry SAM expression (7) is interpreted as the orbital
angular momentum of light, and putting $\hbar$ by hand in (9) it is claimed that
the light beam carries OAM of $l\hbar$ per photon.

Generalization to the circularly polarized light is obtained assuming
\begin{equation}
{\bf A} =(\alpha \hat {\bf x} +\beta \hat {\bf y}) u e^{i(kz-\omega t)}
\end{equation}
and calculating the electric and magnetic field vectors in the paraxial approximation.
The Poynting vector is found to be
\begin{equation}
{\bf S}= {\bf S}^{(1)} +{\bf S} ^{(2)}
\end{equation}
\begin{equation}
{\bf S}^{(2)} =\frac {i\omega \epsilon _0}{2} (\alpha \beta ^*-\beta \alpha ^*) \nabla
{(uu^*)} \times \hat {\bf {z}}
\end{equation}
Expression (12) is interpreted as a spin dependent part of the Poynting vector since
the complex quantities $\alpha ,\beta$ determine the polarization of light. Denoting
$i(\alpha \beta ^* -\beta \alpha ^*)$ by $\sigma$ that takes values $\pm 1$ for
left/right circular polarizations the z-component of the angular momentum density is
given by the sum of $j_z^{(1)}$ i. e. Eq.(7) and
\begin{equation}
j_z^{(2)} =-\frac {\epsilon _0}{2} \omega r \sigma \frac {\partial {|u|^2}}{\partial r}
\end{equation}
Integrating angular momentum density and energy density over the transverse plane the ratio of
the AM to energy per unit length is obtained to be
\begin{equation}
\frac {J_z}{W} =\frac {l+\sigma}{\omega}
\end{equation}
This completes the summary of the main results of \cite{6,16}. It is pertinent to make
certain remarks in the following.

1):- The Hermite-Gaussian (HG) and LG modes in lasers had been extensively studied \cite{12,19},
however the recognotion that the LG modes carry OAM in 1992 \cite{6} stimulated enormous
activity in this field. Historically Poynting in 1909 \cite{22} using mechanical analogy 
associated angular momentum transfer to the optical media from the circularly polarized
light that was first measured by Beth \cite{23}. In the usual interpretation this experiment
is believed to validate the concept of intrinsic spin of photon. Questioning the notion
that the spin of electron is some intrinsic quantum property having no classically
understandable picture, Ohanian \cite{24} argued that the spin could be related with the
circulating energy flow; a nice discussion on the AM of the electromagnetic field is 
also given by him. Both SAM and OAM are calculated for a quasi-plane wave. An illuminating
discussion on the AM of light and the limitations of plane wave approximation can be found 
in Section 2.7 and Chapter 9 of \cite{14}. Interestingly Problems 6.11 and 6.12 in Jackson's
book \cite{1} capture the essential intricacies of the AM of radiation; note that the
expression for the electric field in Problem 6.11 is same as the one obtained using 
Eq. (10) above in the Allen et al work. The identification of the ratio of the AM to 
energy in Problem 6.12 is left ambiguous: whether spin or orbital. This ambiguity
reappears in Section 16.8 where multipole expansion for a vector plane wave is presented
and the value of $m=\pm 1$ for a circularly polarized wave is interpreted as $\pm 1$ unit
of AM per photon along the direction of the wave propagation. Here $m$ is the index 
specifying the $\phi$-dependence of the spherical harmonics $Y_{lm}(\theta ,\phi)$ of order
$(l,m)$. In the paraxial beam Eq. (12) does indicate circulating energy flow i. e. the
presence of $\nabla |u|^2$ term in agreement with the suggestion of Ohanian \cite{24}. 
Though the separation of OAM and SAM is achieved in Eq. (14) a thorough analysis of the
role of scalar function in the SAM seems necessary.

2):- Barnett and Allen \cite{25} raise the question whether the AM for the LG modes 
calculated in \cite{6} is an artefact of the paraxial approximation. Specially the
separation of the total AM into the spin and the orbital parts is discussed. In a later
paper \cite{26} it is shown that the AM flux for a light beam could be separated into
gauge invariant spin and orbital parts without making paraxial approximation. Let us
remember that in the covariant formulation of the electrodynamics \cite{27} conservation 
laws follow from the invariance of the action functional, and the issue of gauge invariance
arises with full complexity. The canonical energy-momentum tensor obtained as a Noether
current from the infinitesimal coordinate transformation is not gauge invariant and also 
not symmetric. Due to the second feature the angular momentum third rank tensor constructed
from the canonical energy-momentum tensor is found to be not a conserved quantity. Adding a
gauge invariant divergenceless, so called spin energy tensor, to the canonical tensor, the
symmetric traceless and gauge invariant energy-momentum tensor is obtained; the 
corresponding angular momentum is conserved. Therefore, the subtle question \cite{28} is 
that of the manifest Lorentz covariance and gauge invariance. Fixing a gauge, for example, the
radiation gauge in optics breaks the manifest Lorentz covariance. Once we assume the radiation
gauge the intricacies of the time-like and longitudinal field excitations (photons) faced in
quantum electrodynamics disappear. It is possible to associate physical observables to spin
and orbital parts of the AM in such a quantum theory, see e. g. \cite{29}. 

Regarding the optical angular momentum flux discussed in \cite{26} remarks in \cite{30}
set the issue in proper context. The main point is that so long as the surface integral
of a divergence term can be made to vanish the angular momentum flux given in \cite{26}
could be useful as a physically observable quantity. In some nontrivial cases the surface
term may give rise to holonomy akin to Aharonov-Bohm effect; prior to the considerations
on the OAM of LG modes \cite{6,25,26} angular momentum holonomy, for both OAM and SAM, was postulated to be the physical mechanism for the geometric phases in optics \cite{31} ; see
Eq. (20) in \cite{31} for the importance of the surface integral term.

Thogh transfer of both SAM and OAM of the light beams to small particles has been 
demonstrated experimentally \cite{32} at a single photon level the issue of the
separation of spin and orbital parts of the angular momentum is not a settled one since
spin seems to possess a metric independent topological attribute.

\section{\bf Gauge invariance and energy-momentum current of light}
Pondering over the remarkably nice form (4) of the Poynting vector (inspired by the remarks in \cite{16}) it struck me that ${\bf S}^{(1)}$ is exactly equivalent to the
3-vector of the conserved current $C^\mu$ associated with the gauge invariance of the complex
scalar field \cite{15}. This analogy is discussed in detail as it may have deep significance. First we make few remarks on
gauge theories to avoid any confusion.
The gauge field theories have electrodynamics as a paradigm. In classical electrodynamics the scalar and vector potentials serve the purpose of mahematical tools as the measurable quantities depend only on the electric and magnetic fields which do not determine the potentials uniquely or alternatively the fields are invariant under the gauge transformation of the potentials, for example, the divergence equation ${\bf \nabla .B}=0$ implies that magnetic field derivable from the curl of a vector potential is invariant under the transformation ${\bf A} \rightarrow {\bf A}+\nabla \chi$. In quantum theory the wave function is arbitrary up to a phase factor. The requirement of local $U(1)$ (phase) gauge invariance in the Lagrangian formulation leads naturally to the electromagnetic interaction. In the textbooks complex scalar field model is usually discussed to illustrate this. Here we are not considering the gauge theory of electromagnetic interactions. We are interested in the global gauge transformation defined by
\begin{equation}
\Psi \rightarrow e^{-i\alpha} \Psi ,~~ \Psi ^{\star} \rightarrow e^{i\alpha} \Psi ^{\star}
\end{equation}
for the complex massless scalar field. The action for this field
\begin{equation}
I=\int L d\tau
\end{equation}
is invariant under the gauge transformation, and the corresponding Noether's conserved current is given by
\begin{equation}
C^\mu =i (\Psi ^{\star} \partial ^\mu \Psi -\Psi \partial ^\mu \Psi ^{\star})
\end{equation}
The Lagrangian density for the complex massless scalar field $\Psi$ is given by
\begin{equation}
L=g^{\mu \nu} \partial _\mu \Psi \partial _\nu \Psi ^\star
\end{equation}
Here Greek indices run from 0 to 3 ; $\alpha$ is a real
gauge parameter; 4-dimensional volume element $d\tau =\sqrt {-g} d^4 x$ where $g$ is
the determinant of the metric tensor $g_{\mu \nu}$. In the flat spacetime $\sqrt {-g} =1$
and $d\tau =c~dt~dV$ where $dV$ is 3-dimensional volume element; $\partial ^\mu =(\frac
{\partial}{c\partial t} ,~ -\nabla)$, and $C^\mu =(C^0,~{\bf C})$.

We make a radical proposition: the scalar field $\Psi$ represents the light
beams and the conservation law for $C^\mu$
\begin{equation}
\partial _\mu C^\mu =0
\end{equation}
is analogous to the continuity equation satisfied by the Poynting vector
\begin{equation}
\nabla .{\bf S}~+~\frac {\partial w}{\partial t} =0
\end{equation}
Recall that $w$ is the energy density of the electromagnetic field. Thus we make the identification
${\bf C}~\rightarrow ~{\bf S}$ and $C^0 ~\rightarrow~w/c$.

First we consider its application to the paraxial light beams. Assuming
\begin{equation}
\Psi=u~e^{i(kz-\omega t)}
\end{equation}
from Eq.(17) a simple calculation shows that
\begin{equation}
{\bf S}^{(1)}~=~\frac {\omega \epsilon _0}{2}~{\bf C}
\end{equation}
Evaluating the time-component
\begin{equation}
C^0 ~ = \frac {2 \omega}{c} |u|^2
\end{equation}
and multiplying it by the factor ${\omega \epsilon _0}/{2}$ we get the energy density (8).
Now the z-component of the 
angular momentum can be easily calculated following \cite{16}. Exact expression for
OAM density (7) and the ratio (9) are found. Since the field
$\Psi$ is scalar with zero spin the angular momentum is necessarily orbital.

It can be argued that this correspondence is an accidental coincidence for the paraxial
beams. To check its validity we apply this formalism to the well known case of
multipole radiation \cite{1}. In this case the scalar
field satisfies scalar wave equation in spherical coordinates $(r, \theta ,\phi)$, for which we assume \cite{1}
\begin{equation}
\Psi = \Psi _0 ~ f_l (kr)~ Y_{l m}(\theta, \phi)~e^{-i \omega t}
\end{equation}
Here $\Psi _0$ is a constant amplitude factor, $f_l$ is the spherical Hankel function which asymptotically (in the radiation zone) behaves as $|f_l|^2$ tending to $\frac {1}{{kr}^2}$. Substituting (24) in Eq. (17) the time component of the gauge current is obtained to be
\begin{equation}
C^0=\frac {2 \omega }{c}~|\Psi _0|^2~|f_l|^2~Y_{lm}^*~Y_{lm}
\end{equation}
and the $\phi$-component is given by
\begin{equation}
C_\phi =\frac {2m}{r sin\theta}~|\Psi _0|^2~|f_l|^2~Y_{lm}^*~Y_{lm}
\end{equation}
The energy of the radiation field in a spherical shell bounded by $r$ and $r+dr$ is calculated
from (25)
\begin{equation}
dw=\frac {2 \omega }{c}~|\Psi _0|^2~|f_l|^2 r^2 dr~\int Y_{lm}^*~Y_{lm} sin \theta 
d \theta d \phi
\end{equation}
The orthonormality of the spherical harmonics give the value of the integral 
equal to 1. Using the asymptotic value of the Hankel function we get
\begin{equation}
\frac {dw}{dr}=\frac {2 \omega}{k^2}~|\Psi_0|^2
\end{equation}
The z-component of the angular momentum in the spherical shell can be calculated from
$r sin \theta C_\phi$ using the asymptotic value of $|f_l|^2$
\begin{equation}
\frac {d J_z}{dr}=\frac {2m}{k^2}~|\Psi_0|^2
\end{equation}
It can be checked that the ratio of the AM to energy is $m/\omega$ , and coincides exactly with that given by 
Eq.(16.66) in \cite{1}. Arguments in \cite{1} based on the quantum
mechanical angular momentum operator, for example, z-component to be $-i\frac {\partial}{
\partial \phi}$ indicate the interpretation of the ratio $\frac {m}{\omega}=\frac {m\hbar}{\hbar \omega}$
in terms of $m\hbar$ units of AM per photon of energy $\hbar \omega$ for the multipole
radiation of order $(l,m)$. Though such an analogy is quite often made in the literature
great caution must be exercised to acribe physical reality to photon or to relate it
with quatum theory. An important point overlooked in most discussions is the obscure nature
of rotational energy of photon: for arbitrary OAM of $l\hbar$ units and spin $\pm \hbar$
the energy per photon is still $\hbar \omega$.

\section{\bf Rotating Light}

Frequency shifts arising from the cyclic polarization changes due to the rotating waveplates
have been related with evolving Pancharatnam phase \cite{33}, however a simple energy exchange
mechanism between the light and the waveplates is shown to explain the observed frequency
shift \cite{34}. The frequency shifts for rotating paraxial beams are also discussed in
the literature \cite{16}. The role of angular momentum in geometric phases in the light
of recent reports is underlined in \cite{35}. The issue of the angular momentum associated
with rotation of the light has received attention quite recently, and there seem to be controversial theoretical results \cite{2,7,8,10}.

Interaction of a monochromatic wave with moving or rotating media would in general result
changes in its frequency; the problem is that of developing an appropriate theory. In
\cite{4} two points are made. A time-dependent generalization of the paraxial Eq.(1) is
proposed
\begin{equation}
\nabla {_t}{^2} u=-2ik(\frac {\partial}{\partial z}~ +\frac {\partial}{c \partial t})u
\end{equation}
And, for a cylindrical lens rotating at constant frequency $\Omega$ it is argued that the 
output beam possesses all frequencies $\omega+2n\Omega,~n=-\infty~to~+\infty$. In \cite{7}
the transverse electric field is expressed in terms of a vector mode function to
incorporate polarization, and the field is assumed to be a superposition of propagating
waves with the phases $\omega _n (t-z/c)$. For a uniform polarization beam rotating with
constant frequency $\Omega$ polychromatic wave expansion for the scalar field is assumed
to be
\begin{equation}
\sum_l u_l (r,z)~e^{il\phi} e^{-i(\omega +l\Omega)(t-z/c)}
\end{equation}
Specifically for two modes with indices $\pm l$ and equal amplitudes $u_l =u_{-l} =u$
the scalar field is
\begin{equation}
2u(r,z)~cos[l(\phi-\Omega (t-z/c))] e^{-i\omega (t-z/c)}
\end{equation}
Bekshaev et al. \cite{2} consider rotating beam comprising of LG modes with $l=\pm 1$,
and term it the rotating HG (RHG) beam. If the frequencies are equal such a superposed beam
is equivalent to an HG beam with an edge wavefront dislocation.

Now elementary analysis shows that a general solution of the wave equation
\begin{equation}
\frac {\partial ^2 f}{\partial z^2}-\frac {1}{v^2} \frac {\partial ^2 f}{\partial t^2}=0
\end{equation}
consists of a superposition of the arbitrary functions $f_1 (z-vt)$ and $f_2 (z+vt)$.
One can introduce harmonic waves with frequencies $\omega _n$ and corresponding
propagation constants $k_n$ to expand these functions in terms of polychromatic waves.
For an ideal monochromatic wave, by definition, there is only one frequency. In the case
of rotating beams, there is such a characteristic frequency i. e. $\omega$ for the input
wave. Since the rotation is affected in the transverse plane, for cylindrical system it
would imply the transformation in the azimuthal coordinate $\phi~\rightarrow~\phi -\Omega t$;
however the propagation along z-axis is still determined by the characteristic frequency
$k=\omega /c$. Therefore the argument suggesting general polychromatic expansion for rotating
beam seems doubtful. It is also surprising that a vast literature on the electromagnetic fields
in the rotating media  \cite{18, 36} has remained unnoticed in this connection. We present a short
review in Sec.VI.

We analyze rotating light using the covariant scalar wave equation
\begin{equation}
\frac {1}{\sqrt {-g}}~\partial _\mu (\sqrt {-g} g^{\mu \nu} \partial _\nu \Psi)=0
\end{equation}
Since the rotation frequency is very small it suffices to employ the so called Galilean
rotation transformation
\begin{equation}
r^\prime =r, ~z^\prime=z, ~t^\prime =t, ~\phi ^\prime=\phi-\Omega t
\end{equation}
Here $(r,\phi,z,t)$ define the inertial system and primed coordinates refer to the uniformly
rotating system. Note that (35) breaks the relativistic covariance; there are subtle issues
of the laboratory, the corotating, and the instantaneous frames of reference \cite{17}.
There does exist Trocheris-Takeno transformation \cite{37} which is known to be relativistically
covariant and has been used to study electromagnetism in rotating media \cite{38}. The
line element expressed in terms of the unprimed coordinates gives rise to the following
nonvanishing metric tensor components
\begin{equation}
g_{00} =c^2 (1-\beta ^2 ),~ g_{11}=-1, ~g_{22}=-r^2 ,~g_{33}=-1,~ g_{02}=g_{20}=\beta r c
\end{equation}
where $\beta =\Omega r/c$. Determinant of the metric is $\sqrt {-g}=rc$, and the
contravariant tensor $g^{\mu \nu}$ has the components
\begin{equation}
g^{00}=1/c^2 ,~g^{11}=-1, ~g^{22}=-(1-\beta ^2)/r^2, ~g^{33}=-1, ~g^{02}=g^{20}=\beta /cr
\end{equation}
The wave equation (34) assumes the form
\begin{equation}
[\frac {1}{c^2} \frac {\partial ^2}{\partial t^2}~-(\frac {\partial ^2}{\partial r^2}~+
\frac {1}{r} \frac {\partial}{\partial r})~-\frac {(1-\beta ^2)}{r^2} \frac {\partial ^2}
{\partial \phi ^2}~-\frac {\partial ^2}{\partial z^2}~+\frac {2\beta}{cr} \frac {\partial ^2}
{\partial t \partial \phi} ]\Psi=0
\end{equation}
Neglecting $\beta ^2$ term it can be seen that this equation is identical with the cylindrical wave equation except the last term. Following the usual prescription let
\begin{equation}
\Psi~=~u(r,z)e^{i(-\omega ^\prime t+kz+l\phi)}
\end{equation}
Assuming that the paraxial equation is satisfied we get the following dispersion relation from (38)
\begin{equation}
{\omega ^\prime}^2 -2l \Omega \omega ^\prime -k^2 c^2 =0
\end{equation}
There are two possibilities: 1) the propagation constant $k$ along z-axis defines a 
characteristic frequency of the incoming wave such that $\omega =kc$; the frequency 
$\omega ^\prime$
of the output beam gets shifted, and 2) the propagation constant is shifted by $l\Omega /c$;
then the frequency is unaffected by the rotation of the media. Assuming the first possibility
we find that unlike the form (30) of ref. 7, also used in \cite{2} the z-dependent phase factor
in Eq.(39) is independent of $l\Omega$.

Consider the rotating beam comprising of two modes with indices $\pm l$, and having
$u_l =u_{-l} =u$. The scalar field $\Psi$ is a superposition of the two modes
\begin{equation}
\Psi = u[e^{-il\Omega t +il\phi} +e^{il\Omega t -il\phi}]~e^{-i\omega t+ikz}
\end{equation}
Expression (17) for the Noether current gives the energy and linear momentum densities. The
metric (36) defining the system implies that the time and azimuthal components of contravariant and covariant current differ. First we calculate the components of (17) for the ordinary 
derivative operators $\frac {\partial}{\partial t} ,~\frac {\partial }{\partial \phi}$.
\begin{equation}
\bar {C_0} = 8\omega |u|^2 cos^2 l(\phi-\Omega t)
\end{equation}
\begin{equation}
\bar {C_\phi} =0
\end{equation}
Compare Eq.(42) with the energy density obtained for similar case in \cite{7}
\begin{equation}
w=8\epsilon _0 |u|^2 cos^2 [l(\phi - \Omega (t-z/c))]
\end{equation}
Since $k$ does not change with $\Omega$ in our analysis it would appear that $\bar {C_0}$ could
be identified with the energy density as proposed in the preceding section. However OAM
density calculated in \cite{7} is nonzero while (43) would give its value zero. Let us
examine the problem calculating the contravariant current components. Note that for a scalar
field, say, f the ordinary derivative $\partial _\mu f$ is a covariant vector, and the contravariant vector is given by $g^{\mu \nu} \partial _\nu f$. For the metric (37) we get
\begin{equation}
\partial ^0 =g^{00} \frac {\partial}{\partial t} + g^{02} \frac {\partial}{\partial \phi}
\end{equation}
\begin{equation}
\partial ^2 =g^{20} \frac {\partial}{\partial t} + g^{22} \frac {\partial}{\partial \phi}
\end{equation}
while r and z components do not change. From (17) using (45) and (46) we finally get
\begin{equation}
C_{rot} ^0 =\bar {C_0} /c^2
\end{equation}
\begin{equation}
C_{rot} ^\phi =\Omega \bar {C_0} /c^2
\end{equation}
We have to keep in mind that in tensor analysis we define physical quantities. Here neglecting
$\beta ^2$ (48) has to be multiplied by $r$ that corresponds to momentum density. The OAM density
is obtained to be
\begin{equation}
j^z _{rot} =r^2 \Omega \bar {C_0} /c^2
\end{equation}
Taking $\bar {C_0}$ as energy density, Eq. (49) could be interpreted as AM of a rigid body 
rotation with angular velocity $\Omega$. However earlier cited works \cite{2, 7, 8} assert that
there is no analogy with the rigid body rotation; this contradicts our result. To clarify the issue
in the next section we present a detailed discussion.

\section{\bf OAM Paradox}

The result obtained in the preceding section is significant: there is no paradoxical feature
that was found in the earlier studies \cite{2,7} on the OAM of rotating light beams; at the
same time the circulating energy pattern described by Eq.(42) agrees with the observations
as interpreted in \cite{2}. To delineate the physical mechanism responsible for the rotating
light we consider a single mode with index $l$ represented by
\begin{equation}
\Psi= u e^{-i \Omega t+il\phi -i\omega t +ikz}
\end{equation}
Using ordinary derivatives (17) gives
\begin{equation}
C^l _0 = 2 |u|^2 (\omega +l \Omega)
\end{equation}
\begin{equation}
C^l _\phi = 2l |u|^2
\end{equation}
Physical azimuthal component is $C^l _\phi /r$, and the OAM density to energy density ratio is derived to be
\begin{equation}
R=\frac {l}{\omega+l\Omega}
\end{equation}
Further the contravariant components are evaluated to be
\begin{equation}
C^{0l} =\frac {2|u|^2 \omega}{c^2}
\end{equation}
\begin{equation}
C^{\phi l} =\frac {2l|u|^2}{r^2}~+~\frac {2\Omega \omega |u|^2}{c^2}
\end{equation}
Physical quantity from (55) is defined multiplying it by $r$, and the OAM density is
obtained to be
\begin{equation}
j^{zl} =2|u|^2 [l+~\frac {\Omega \omega r^2}{c^2}]
\end{equation}
Eq. (56) shows that interpreting $C^{0l}$ as energy density the OAM density comprises
of two parts: the first term on the right side is OAM density corresponding to the nonrotating beam with the azimuthal index $l$, and the second term represents the rigid rotation of
the beam with angular velocity $\Omega$.

Doing a naive calculation adding energy density and OAM density for the modes with
opposite indices $\pm l$ we get
\begin{equation}
C_0 (rot)= C_0 ^l~+~C_0 ^{-l}=4 |u|^2 \omega
\end{equation}
\begin{equation}
j_z (rot) =0
\end{equation}
Apart from the missing cosine squared term in (57) these are consistent with Eqs. (42) and
(43). Suppose we calculate the ratio from (53) for the mode $-l$ and add the two then we find
\begin{equation}
R(rot) = \frac {l}{\omega +l\Omega}~-~\frac {l}{\omega -l\Omega}=-\frac {2l^2 \Omega}
{\omega ^2 -l^2 \Omega ^2}
\end{equation}
This is an startling result: Eq.(59) bears close resemblance with the main paradoxical feature
i. e. Eq.(39) in \cite{7} and Eq.(21) in \cite{2}. In fact, rewriting (59) in the form
\begin{equation}
R(rot)=\frac {l\hbar}{(\omega +l\Omega)\hbar}~-~\frac {l\hbar}{(\omega -l\Omega)\hbar}
\end{equation}
it would seem that following \cite{2,7} invoking rotating photons with OAM of $\pm l \hbar$
and energy $(\omega \pm l\Omega)\hbar$ a physical interpretation of this paradoxical result
could be envisaged. However in our analysis the derivation of (59) is incorrect: it is
an artefact of the derivation. Physically meaningful would be the sum of contravariant
quantities i. e. using Eq.(56) we find that the OAM arising as a rigid body rotation
is consistent with Eq.(49).

The distinction between co-, contra-, and physical vectors and their interrelationship
may appear intricate or at times annoying; however this is the limitation imposed by the
metric based formalism. One has to be careful about it, for example, in the standard
3 dimensional space in curvilinear coordinate system. Even in flat spacetime geometry
treating the time coordinate as $x^0 =t$ or $x^0 =ct$ results into different quantities;
in Sec.II it is assumed that $x^0 =ct$, thus $\sqrt {-g} =1$ whereas in Eq.(35) $x^0 =t$
leading to the appearance of the velocity in odd form subsequently.

To sum up: a new perspective on the OAM of rotating light emerges in the covariant scalar
field approach, and though the paradoxical result akin to the one reported in the literature
could be obtained it is obviously wrong in the present formalism.

\section{\bf Electromagnetic fields in noninertial systems}

The formal structure of the Maxwell equation is relativistically covariant; mathematical
aspects are eloquently discussed in Eddington's book \cite{39}. A consistent application
of the covariant formalism to the field equations sheds light on the constitutive relations
in the rotating media, and as argued in \cite{36} offers a satisfactory resolution of the
Schiff paradox \cite{40}. However there do exist subtle issues \cite{41}. In the limited context
of the present paper a short commentary on the radiation in rotating system seems useful.
An important example is that of Sagnac effect known since 1913 and reviewed in \cite{18}.
To avoid any confusion first we recall that the Sagnac effect occurs in the rotating 
interferometer experiments. Light emanating from a source is split into two beams which
are made to circulate in opposite directions along a closed loop, and recombined to give
the interference pattern. If the whole system is set into rotation than the fringe shift
relative to the stationary system is observed. The whole system consists of the interferometer,
light source and detector, and the medium. Variants of this experiment also show fringe shifts:
the medium is stationary while the interferometer rotates, and the medium is rotated keeping
the interferometer stationary. Theory of this phenomenon is still controversial \cite{18,42}.

Assuming cylindrical system light propagation is along the circular path in the Sagnac effect
in contrast to the rotating beams which travel along the z-axis. However in spite of this difference the consideration of
the electromagnetic fields in noninertial frame is crucial and common in both cases. Some
of the interesting papers are those of Heer \cite{43}, Anderson and Ryon \cite{44}, and
Post \cite{45}. Heer calculates resonant frequency of a cavity in the rotating system. The
constitutive relations used by him are criticized in \cite{44}. A noteworthy result for
the cylindrical cavity \cite{43} is the axial modes splitting with frequency
\begin{equation}
\omega _m =\omega _m ^0~\pm ~m\Omega
\end{equation}
where $m$ is the azimuthal index (other indices are suppressed) and $\Omega$ is the rotation 
frequency along z-axis. Multiplying on both sides by $\hbar$ and introducing the OAM of photon 
$J_z$ this equation is rewritten as
\begin{equation}
\hbar \omega =\hbar \omega ^0~ +J_z \Omega
\end{equation}
Heer calls it Coriolis-Zeeman effect for the photon. Anderson and Ryon develop a generally covariant formalism for an arbitrarily moving medium. In the special case of uniform rotation
the frequency shift found by them does not depend on the relative permeability of the medium
contradicting the result of Post \cite{18}. In a later article Post \cite{45} has analyzed the
basic questions related with the problem, and argued that though the physical interpretation
of the first order effects in noninertial systems remain unsettled, the experiments on unipolar
induction (Kennard-Pegram effect) with different media could throw light on the issue of
the constitutive relations. This brief review with a limited and incomplete citations shows
that there are open problems in the subject of electromagnetism in rotating media \cite{18, 38, 45}.

Covariant scalar field approach presented in this article is a step forward in this direction, however for an unambiguous resolution of the problem a consistent covariant formulation of
the electromagnetic fields in the rotating media is suggested. Further credence to this suggestion
is obtained analyzing the calculation in \cite{8}. Authors use quantum mechanical analogy
and generalize the formula of \cite{9} for quasi-monochromatic light. The main change is contained in the time derivative of a vector reproduced below
\begin{equation}
\frac {\partial {\bf A}}{\partial t} = -i\omega {\bf A} +\vec{\Omega} \times {\bf A}
-\Omega \frac {\partial {\bf A}}{\partial \phi} 
\end{equation}
According to the authors fast time dependence is through the factor $e^{-i\omega t}$, and
angular velocity vector is $\vec{\Omega} = (0,0,\Omega)$. First two terms on the right
of (63) are understandable, however the last term is peculiar. To obtain (63) physical
arguments are put forward in the paragraph preceding this i. e. Eq.(7) in \cite{8}. We examine it in covariant formalism.

The definition of a covariant derivative of a vector \cite{39} is
\begin{equation}
D_\nu A_\mu =\partial _\nu A_\mu -~ \Gamma ^\alpha _{~\mu\nu} A_\alpha
\end{equation}
Here $\Gamma ^\alpha _{~\mu\nu}$ is the Christoffel symbol. Setting the indices $\nu =0;\mu =1,2,3$
in Eq.(64) we get the time component of the covariant derivative for the spatial components of
the vector $A_\mu$. In the special case of uniform rotation with the metric (36) Eq.(64) gives 
expression (63) without the last term. What is the origin of the last term? Besides the covariant derivative of a covariant vector, there are three more quantities \cite{39} : covariant
derivative of $A^\mu$, and contravariant derivatives of $A_\mu$ and $A^\mu$. Let us calculate
the contravariant derivative. We have to use (45) for the operator $D^0$, substitute $g^{02}$
from (37), and evaluate $\Gamma ^\alpha _{~\mu\nu}$. After a little lengthy calculation we
get finally expression (63) if we assume
\begin{equation}
\frac {\partial {\bf A}}{\partial t} =-i\omega {\bf A}
\end{equation}
It is remarkable that physical intuition led the authors of \cite{8} to the time component
of a contravariant derivative. The significance of this identification is two-fold: 1) the
approach in \cite{8} is based on quantum mechanical analogy and the generalization of time
derivative by them finally results into the conclusion in agreement with \cite{2,7}. Viewed from
the covariant perspective the derivation of electric and magnetic fields using only Eq.(63) for
time derivatives is incorrect since these fields are not vectors but the components of a second rank tensor $F^{\mu \nu}$. Thus the final result in \cite{8} is questionable. And, 2) For a
conclusive resolution of the issue it becomes imperative that a rigorous generally relativistic
covariant treatment of the rotating media be incorporated. Though constitutive relations will
be fairly involved and the analysis quite complicated, since fundamental issues are involved
such a study will be important.

\section{\bf Discussion and Conclusion}

Recent theoretical studies show that rotating light beams possess angular momentum (OAM
and SAM) having paradoxical attributes. Attempts have been made to explain such unexpected
features, however the physical interpretation still lacks clarity. In this paper the OAM of
rotating light is examined afresh. An important ingredient in our approach is the recognition
that Poynting vector continuity equation in the case of paraxial beams is equivalent to the
Noether gauge current conservation law for a complex scalar field \cite{46}. Postulating scalar
field $\Psi$ to represent light without relating it to the electromagnetic fields, a covariant
formulation is developed. This allows us to treat uniform rotation adopting the general relativistic framework. The OAM of rotating light is easily calculated using the gauge current
$C^\mu$. It is shown that there is no paradox if the distinction between co-, contra-, and 
physical vectors is taken care of. A significant result is that the OAM of the rotating beam
comprising of two modes with opposite azimuthal indices is consistent with the picture of a
rigid body rotation. This contradicts the previous literature \cite{2,7,8}. The matter is
further elucidated in Sec.V where it is shown that one can 'manufacture' a formal
paradoxical result, however it is wrong in the present covariant approach.

The interpretation of $C^\mu$ as energy-momentum vector is motivated by the nice form of the
Poynting vector (4) for the paraxial light. Though it has been justified showing its
applicability to the calculation of the angular momentum of the multipole radiation, there
remain some basic questions to be addressed. In field theory the energy-momentum vector is 
obtained setting the index $\nu =0$ in the energy-momentum tensor $T^{\mu \nu}$. On the other 
hand, the gauge current is associated with the charge of the field such that local gauge
invariance naturally leads to the electromagnetic interaction via gauge covariant derivative
\cite{15, 47}. Interestingly for the special case of the complex scalar field we find an
important result: the tensor $T^{\mu 0}$ for $\Psi$ gives the momentum vector formally
equivalent to ${\bf C}$ assuming $e^{-i\omega t}$ time-dependence for $\Psi$. 
Note that in flat spacetime
$g^{\mu 0}$ vanishes for $\mu =1,2,3$ and the expression $\partial ^\mu \Psi \partial ^\nu \Psi ^*
+\partial ^\nu \Psi \partial ^\mu \Psi ^*$ in $T^{\mu \nu}$ reduces to the form of ${\bf C}$.
However the time-component $T^{00}$ disagrees with $C^0$. We may think of enlarging the group of symmetries including conformal invariance; one requires energy-momentum tensor to be traceless
in that case. One can add a term to $T^{\mu \nu}$ keeping intact the covariant divergence law
\cite{47} and derive a new traceless tensor; this can also be done adding a surface term to the
Lagrangian (18). The role of gauge and conformal symmetries envisaged here would have wider
ramification for the the gauge theories \cite{48} and deserves further investigation.

There are two major shortcomings of the present work. First, we have not considered spin angular momentum. In the paraxial approximation assuming Eq.(10) for the vector potential the spin-dependent Poynting vector (12) is derived, see \cite{16}. Nienhuis \cite{7} has generalized it
to the rotating polarization beams and considers polychromatic waves. An intriguing discussion
in this paper relates with inhomogeneous linear polarization where entangled photons are suggested. In a subsequent work \cite{10} introducing plausible quantized rotating mode
operators a quantum theory of rotating light is developed. In Sec.V authors calculate SAM
for single photon which is, similar to OAM, found to be counterintuitive. Obviously the
scalar field theory is inadequate to address the question of spin of photon, and it is as yet
not clear how to generalize our work for this purpose. We may, however point out that spin is
intrinsic and possesses the characteristic of metric-independence \cite{49}; the results on
single photon found in \cite{10} cannot be accepted without reservation.

Second limitation arises because we do not consider electromagnetic fields: though scalar wave theory does explain a number of properties of light, we cannot be realistic unless electromagnetic
fields are considered. To motivate general relativistic treatment of the full Maxwell field theory
a short review is presented in the preceding section on the electromagnetism in noninertial
systems. Further justification for this kind of approach is sought re-analyzing the theory
given in \cite{8}. The main modification in \cite{8} is based on the definition of the time-derivative of a vector proposed by the authors that is subsequently used in Maxwell equation
to calculate electric and magnetic fields. In the present paper we show that the new time-derivative is essentially the time-component of the contra-variant derivative of a vector
in the metric space defined by Eq.(36). It is logical to argue that general relativistic
formalism is necessary to resolve the controversial issues; not only this perhaps the
advances in rotating light experiments may throw light on some outstanding problems in the
electromagnetism of rotating media (Sec.VI); an interesting and controversial problem is that of
the scattering of radiation from a rotating cylinder discussed by Hillion \cite{38}.

In conclusion, equivalence of gauge current for a complex scalar field and energy-momentum vector
for paraxial light is shown; a covariant scalar field theory in rotating system is presented
to address the issue of the OAM of rotating light, and it is shown that there does not arise any
paradox highlighted in the recent literature.

Acknowledgements

I gratefully acknowledge correspondence with Prof. L. Allen, Dr. A. Yu. Bekshaev, and Prof. G. Nienhuis.
Library facility of Banaras Hindu University is acknowledged.

\end{document}